\documentclass[fleqn,usenatbib,onecolumn]{mnras}
\usepackage{hyperref}

\usepackage[T1]{fontenc}

\DeclareRobustCommand{\VAN}[3]{#2}
\let\VANthebibliography\thebibliography
\def\thebibliography{\DeclareRobustCommand{\VAN}[3]{##3}\VANthebibliography}

\usepackage{graphicx}	
\usepackage{amsmath}	
\usepackage{amssymb}	
\usepackage{multicol}
\usepackage{mathptmx}
\usepackage{txfonts}

\title[Polarized hotspots in Sgr A*'s accretion flow]{Polarized signatures of adiabatically expanding hotspots in Sgr A*'s accretion flow}

\author[Michail et al. (2023)]{
Joseph M. Michail,$^{1,2}$\thanks{michail@u.northwestern.edu}
Farhad Yusef-Zadeh$^{1,2}$,
Mark Wardle,$^{3}$ and
Devaky Kunneriath$^{4}$
\\
$^{1}$Department of Physics and Astronomy, Northwestern University, 2145 Sheridan Rd, Evanston, IL 60208, USA\\
$^{2}$Center for Interdisciplinary Exploration and Research in Astrophysics (CIERA), Northwestern University, 1800 Sherman Ave, Evanston, IL 60201, USA \\
$^{3}$Research Centre for Astronomy, Astrophysics and Astrophotonics and School of Mathematics and Physical Sciences, Macquarie University, \\Sydney, NSW 2109, Australia\\
$^{4}$National Radio Astronomy Observatory, 520 Edgemont Road, Charlottesville, VA 22903, USA
}

\date{Accepted XXX. Received YYY; in original form ZZZ}

\pubyear{2023}

\begin{document}
\label{firstpage}
\pagerange{\pageref{firstpage}--\pageref{lastpage}}
\maketitle

\begin{abstract}
    We report 235 GHz linear and circular polarization (LP and CP) detections of Sgr A* at levels of $\sim10\%$ and $\sim-1\%$, respectively, using ALMA. We describe the first full-Stokes modeling of an observed submillimeter flare with an adiabatically-expanding synchrotron hotspot using a polarized radiative transfer prescription. Augmented with a simple full-Stokes model for the quiescent emission, we jointly characterize properties of both the quiescent and variable components by simultaneously fitting all four Stokes parameter light curves. The hotspot has magnetic field strength $71$ G, radius $0.75$ Schwarzschild radii, and expands at speed $0.013$c assuming magnetic equipartition. The magnetic field's position angle projected in the plane-of-sky is $\approx55^\circ$ East of North, which previous analyses reveal as the accretion flow's angular momentum axis and further supports Sgr A* hosting a magnetically-arrested disk. The magnetic field is oriented approximately perpendicular to the line of sight, which suggests repolarization as the cause of the high circular-to-linear polarization ratio observed at radio frequencies. We additionally recover several properties of the quiescent emission, consistent with previous analyses of the accretion flow, such as a rotation measure $\approx-4.22\times10^{5}$ rad m$^{-2}$. Our findings provide critical constraints for interpreting and mitigating the polarized variable emission in future Event Horizon Telescope images of Sgr A*.
\end{abstract}




\section{Introduction}
    Sagittarius A* (Sgr A*) is the $(4.152\pm0.014)\times10^{6}~\rm{M}_\odot$ supermassive black hole located at the Galactic Center at a distance of $8178$ pc from the Earth \citep{Gravity2019}. Sgr A* is well-known to vary across the electromagnetic spectrum \citep[e.g.,][]{FYZ2011, Neilsen2013, Subroweit2017, Do2019, Witzel2021, Wielgus2022a}. In the radio and submillimeter (submm), the emission is dominated by two components, both of which originate from the accretion flow: (quasi-)quiescent and variable radiation. The accretion flow is turbulent, which causes variations in its flux on timescales $\gtrsim$ minute. The accretion flow produces an overall net level of quiescent emission on top of low- and high-level amplitude changes owing to its variable nature. The large-amplitude variations are known as ``flares'' and dominate the variable emission. Previous work on Sgr A*'s variability in the radio/submm regimes has focused on total intensity observations. However, these earlier results largely ignore the polarization of Sgr A*, which helps uncover the accretion flow's magnetic properties.
    
    \cite{Bower1999a, Bower1999c, Bower2001} first searched for linear polarization (LP) between 5-112 GHz and found Sgr A* to be linearly unpolarized at these frequencies. However, circular polarization (CP) was detected at 5 and 8 GHz \citep{Bower1999b, Sault1999}. Further observations extended polarimetric wavelength coverage from 1.5 GHz to 230 GHz \citep{Bower2002,Tsuboi2003,Munoz2012}. \cite{Aitken2000} first hinted at intrinsic LP from Sgr A* at 400 GHz; however, their measurements may have been contaminated by the abundance of polarized dust immediately surrounding Sgr A* in the circumnuclear disk \citep[e.g.,][]{Hsieh2018}. 
    
    The first interferometric observations detected LP at 230 and 340 GHz \citep{Bower2003, Marrone2006}. Later observations broadened this range from 86 to 700 GHz \citep{Macquart2006, Liu2016a, Liu2016b}. Presently, Sgr A* is known to be circularly polarized from 1.5-230 GHz and linearly polarized between 86-700 GHz. \cite{Liu2016a,Liu2016b} find Sgr A*'s LP percent to increase from $\sim0.5\%$ at 93 GHz to $\sim8.5\%$ at 500 GHz, which may decrease at higher frequencies. \cite{Munoz2012} compiled CP measurements of Sgr A* from 1.5 GHz to 345 GHz finding levels of $\sim-0.2\%$ to $\sim-1\%$, respectively. While the absolute CP amplitude is known to vary \citep[see][]{Munoz2012}, the \textit{sign} is consistently negative in the radio and submm in all currently published data, which they suggest is caused by a highly coherent magnetic field configuration throughout the accretion flow.
    
    In addition to studying Sgr A*'s long-term polarimetric trends \citep[i.e.,][]{Bower2002, Bower2005, Macquart2006, Munoz2012, Bower2018}, the detection of hourly-timescale variation of LP by \cite{Marrone2006} opened up an additional avenue by which to study the accretion flow via the variable emission. Several models describing the total intensity flaring emission have been proposed, such as jets/outflows \citep{Falcke2000, Brinkerink2015} and adiabatically-expanding synchrotron hotspots embedded in the accretion flow \citep{VDL1966, FYZ2006}. 
    \cite{FYZ2007} first modeled the full-Stokes light curves of these hotspots using an analytic formalism for the transfer of polarized synchrotron radiation through a homogeneous medium \citep{Jones1977a}. Supplementing this simple picture with full-Stokes radiative transfer presents a new opportunity to study the magnetic field configuration in a localized region of the accretion flow. Previously, only the equipartition magnetic field strength could be estimated from this model. However, the observed polarization light curves are regulated by the orientation of the magnetic field relative to the observer. The orientation is a crucial physical parameter that could not previously be determined. \cite{FYZ2007} tested this full-polarization hotspot model at 22 and 43 GHz; however, their analysis was limited as the LP level was low ($\sim0.2-0.8\%$), and the data were noisy. Sgr A* is brighter and more linearly and circularly polarized at submm frequencies, which decrease the overall uncertainty in the polarimetric properties.
     
    In this paper, we present the first full-Stokes modeling of Sgr A*'s submm flaring emission using the adiabatically-expanding hotspot model. This paper is organized as follows. In Section \ref{sec:data}, we discuss the observations and processing of the data and analyze possible systematic issues in the CP products. In Section \ref{sec:analysis}, we describe the models adopted for the quiescent and variable components used to fit the full-Stokes light curves and present the best-fit values. For the first time, we determine the orientation of the hotspot's magnetic field on the plane-of-sky and along the line-of-sight. We find the projected magnetic field to be oriented along the Galactic Plane and approximately perpendicular to the line-of-sight. This has interesting implications for the accretion flow's magnetic field configuration, which we discuss in Section \ref{sec:disc}. Furthermore, in Section \ref{sec:disc}, we discuss the other results in the context of previous analyses in the literature and consider some limitations with our chosen data set. Finally, in Section \ref{sec:summary}, we present a summary of our findings and discuss future work.

\section{Data}\label{sec:data}
    \subsection{Observations and Processing}\label{ssec:observations}
        The Atacama Large Millimeter/submillimeter Array (ALMA) observed Sgr A* on 16 July 2017 in band 6 ($\approx230$ GHz) in full polarization (project ID 2016.A.00037.T). These data, part of a multi-wavelength campaign of Sgr A* concurrent with the Chandra X-ray Observatory and the Spitzer Space Telescope, were taken with the 12-meter array in the C40-5 configuration (the baselines ranged from 17 to 1100 meters). For our analysis, we focus only on the submm data. 
        
        The observation consists of two line and two continuum spectral windows. The two continuum windows are centered on 233 and 235 GHz, each having a bandwidth of 2 GHz with 64 31.25-MHz bandwidth channels. The first spectral line window is centered on SiO (5-4) at $\approx$217 GHz with a 1.875 GHz bandwidth of 1920 0.976-MHz channels. The second spectral line window is centered on $^{13}$CO (2-1) at 220.398 GHz with 1920 0.244-MHz channels for a total bandwidth of $\approx0.47$ GHz. In this analysis, we average over all of the channels per spectral window to obtain four frequency-averaged continuum windows.
        
        Only one of five execution blocks was observed owing to technical issues which occurred during the observation. We used the ALMA pipeline (version 2020.1.0.40) with CASA 6.1.1.15 \citep{McMullin2007} to generate the calibrated data. The following calibrators are used to generate the calibration tables: J1733-1304 (flux), J1517-2422 (bandpass), J1549+0237 (instrumental polarization), and J1744-3116 (phase). The QA2 team designated these data ``semi-pass'' since the parallactic coverage ($\approx46^\circ$) was lower than recommended to determine the instrumental polarization terms ($60^\circ$). Despite this, we were able to calibrate the instrumental polarization. We imaged and phase self-calibrated the data starting with a solution interval of $30$ seconds and stopping at a single integration time ($6.05$ seconds). After phase self-calibration, we flagged any obvious misbehaving baselines or antennas.
        
            \begin{figure*}
            \centering
            \includegraphics[trim = 1.8cm 0cm 0cm 0.0cm, clip, width=0.95\textwidth]{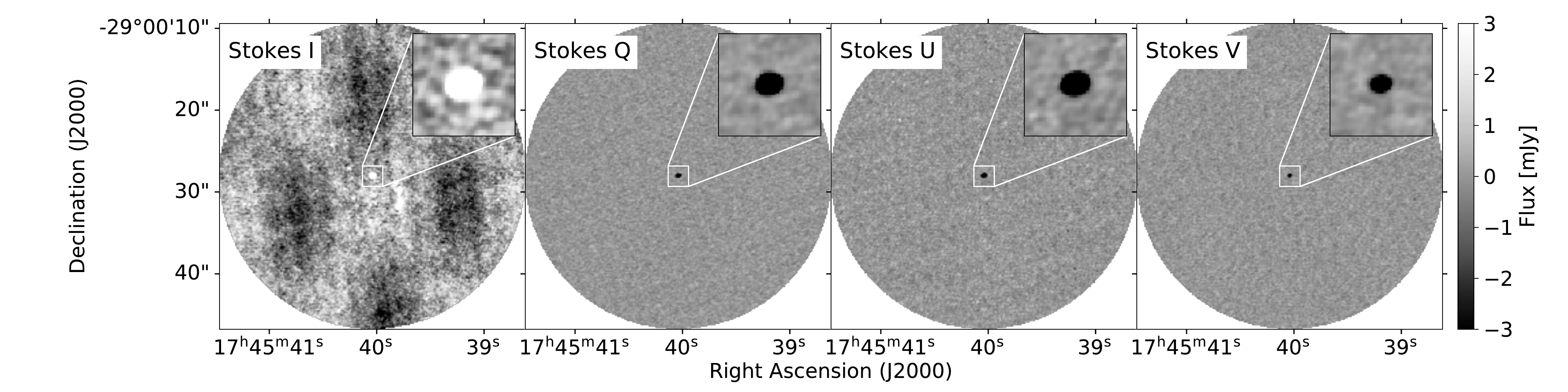}
            \caption{A sample of example images of Sgr A* on 16 July 2017 at 00:28:07 UTC using a 30-second binning time for each Stokes parameter at 233.5 GHz. The full image is $\approx51\arcsec$ per side. We flag pixels below a normalized primary beam limit of 20\%, resulting in an image that is approximately $37\arcsec$ per side. The panels use the same gray scale to show the noise level. The inset is a $2\farcs5\times2\farcs5$ subregion centered on Sgr A*.}
            \label{fig:example_binned}
        \end{figure*}
        
        We developed a CASA script to autonomously determine the full polarization light curves for a point source located at the phase center using \texttt{TCLEAN} and \texttt{IMFIT}. Briefly, the code bins every scan on a single source to a user-defined value for imaging. \texttt{TCLEAN} images the visibilities in all four Stokes parameters for each time bin. \texttt{IMFIT} is used to fit a point source + zero-level offset at the phase center, where Sgr A* is located, in each image and polarization to determine the point source flux density and (statistical) error. We construct the point source light curves using the fitted \texttt{IMFIT} parameters and export them to a text file for further analysis, where we calculate the polarization product light curves (see Appendix \ref{appendix:pol_conv} for our chosen conventions). Since Sgr A* is surrounded by diffuse emission which is not fully resolved out in the observed configuration, this method yields contamination-free light curves without restricting the projected baseline length, which would lower the overall sensitivity.
        
\begin{figure*}
        \centering
        \includegraphics[trim = 0.8cm 1.3cm 2.25cm 0.5cm, clip, width=0.49\textwidth]{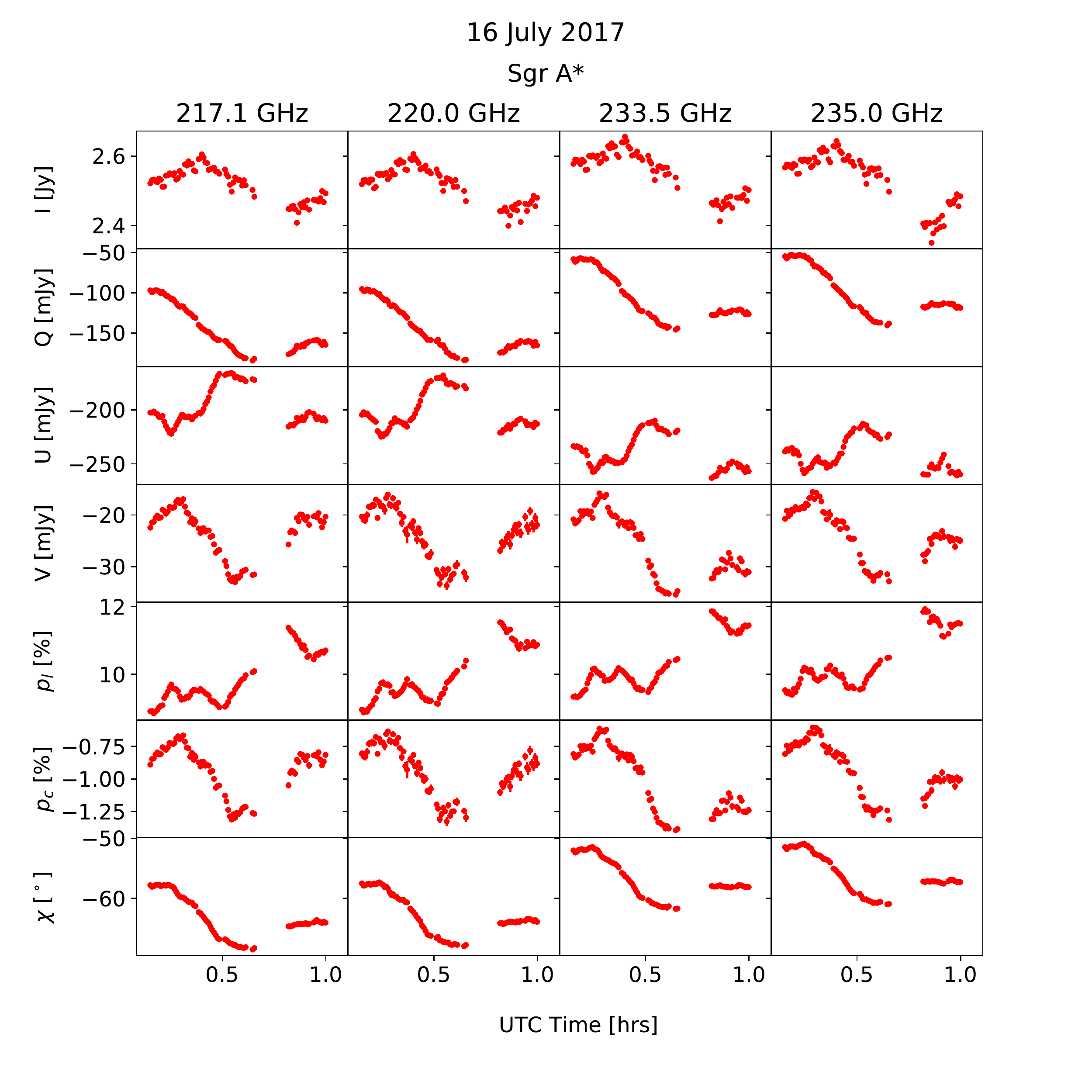}
        \includegraphics[trim = 0.8cm 1.3cm 2.25cm 0.5cm, clip, width=0.49\textwidth]{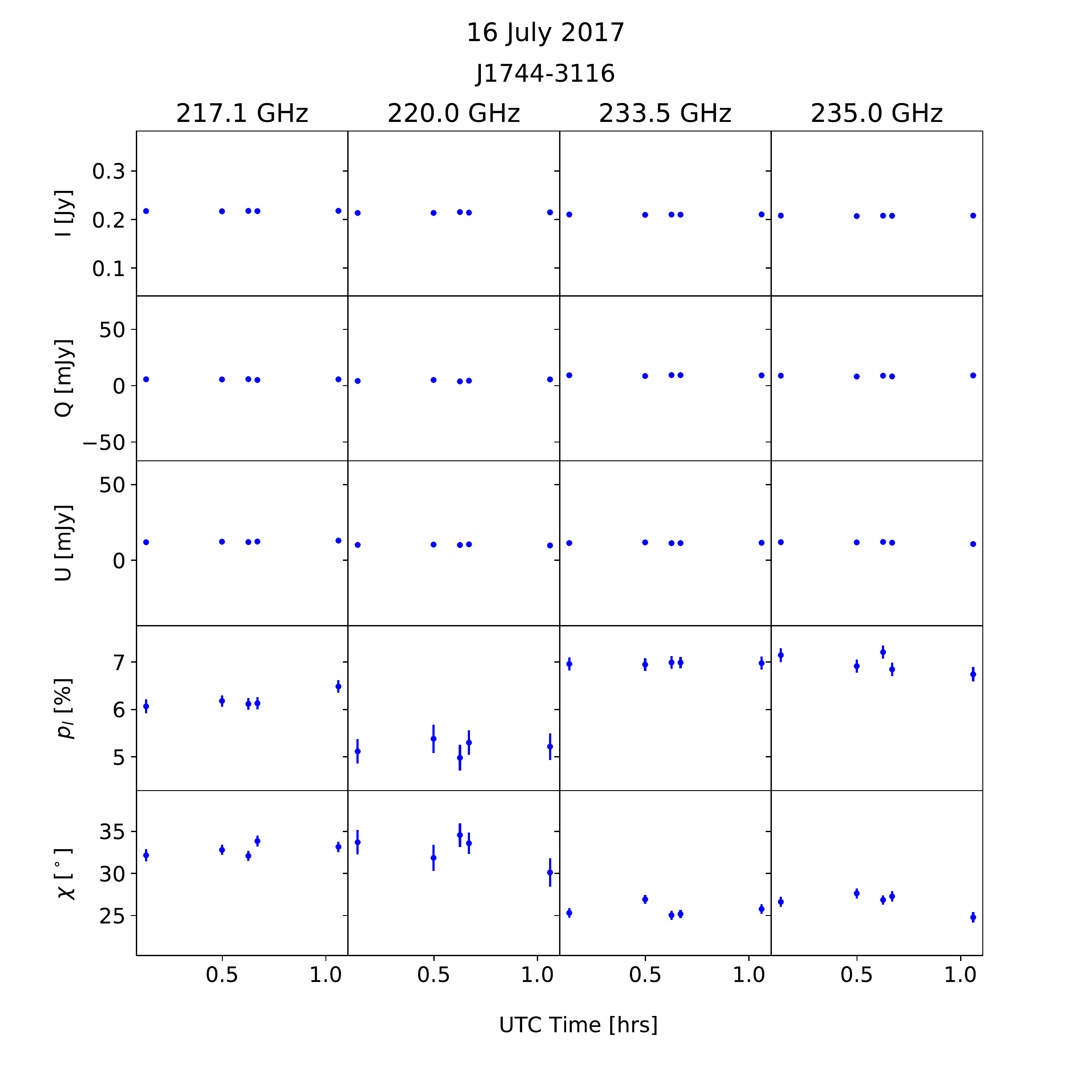}
        \caption{Full Stokes and polarization light curves of Sgr A* (red, left) and the phase calibrator J1744-3116 (blue, right) on 16 July 2017. As J1744-3116 is not circularly polarized (see Section \ref{ssec:circular_pol}), not shown are Stokes V and $p_c$ for this source. Error bars for both sources are shown and are often smaller than the marker size.}
        \label{fig:july_lcs}
    \end{figure*}

        We use 30-second binning in our analysis. Each image is $1024\times1024$ ($\approx51\arcsec\times51\arcsec$) and uses a cell size of $0\farcs05$. We apply the standard 20\% primary beam cut to remove imaging artifacts toward the edge of the image, resulting in a final image size of $\approx37\arcsec\times37\arcsec$. We do not primary-beam correct the image since Sgr A* is at the phase center. We restrict the maximum number of iterations to $1000$ to properly clean any extended emission while not cleaning noise artifacts. We show a sample image of Sgr A* during a 30-second binning time in all four Stokes parameters in Figure \ref{fig:example_binned}. In Figure \ref{fig:july_lcs}, we show the final Stokes I, Q, U, and V, LP percent ($p_{l}$), CP percent ($p_{c}$), and LP angle ($\chi$) light curves used in our analysis. Overall, we find Sgr A* to be linearly and circularly polarized at levels of $\sim10\%$ and $\sim-1\%$, respectively. The definitions of these parameters and their uncertainties are detailed in Appendix \ref{appendix:pol_conv}. We discuss the absence of CP products for J1744-3116 in Section \ref{ssec:circular_pol}.

    \subsection{Verifying Sgr A*'s Circular Polarization Detection}\label{ssec:circular_pol}
            There has been great care taken in previous polarimetric analyses to rule out calibration-error-based CP detections. \cite{Goddi2021} present a detailed description of the issue (see their Appendix G). In short, the polarization calibrator is assumed to have Stokes $V = 0$, which can induce a false (and time-dependent) CP onto the target sources. To check for systematics, we focus on the CP characteristics of J1517-2422 and J1744-3116 following the prescription given in \cite{Munoz2012}. J1517-2422 has a similar declination to Sgr A* ($17^{\rm{h}}45^{\rm{m}}40.04^{\rm{s}},-29^\circ00\arcmin28.17\arcsec$), while J1744-3116 has a comparable right ascension. To check for intrinsic CP for the calibrators, we image each spectral window in Stokes I and V during the entire observing window using the same non-interactive process in Section \ref{ssec:observations}. We obtain a higher sensitivity image to detect CP by imaging the entire observation. We fit a point source to the phase center using \texttt{IMFIT} and report the integrated flux density and statistical error. These results are shown in Table \ref{tab:calibrator_pc}. While \texttt{IMFIT} returns converged flux densities and errors for J1744-3116 in Stokes V, the images do not show circularly polarized emission at or near the phase center. To quantify the $3\sigma$ upper limit on CP, we calculate $3\times$ the Stokes V root-mean-square (RMS) provided by \texttt{IMSTAT}. 
            
            For J1517-2422, we detect a statistically significant $p_c \approx -0.1\%$. Since this source is bright ($>3$ Jy), residual or uncalibrated instrumental polarization terms in V could lead to spurious CP measurements. Despite the unfortunately sparse coverage of this source, we compare our results to those in the literature. \cite{Bower2018} report $p_c\approx0.1\%$ for this source in August 2016 at $\approx240$ GHz, having the same magnitude (but opposite sign) as our result. Following \cite{Goddi2021}, we use the AMAPOLA\footnote{\url{http://www.alma.cl/~skameno/AMAPOLA/}} project, which tracks the flux density and polarization properties of several ALMA calibrators for more nearby observations to July 2017. At 233 GHz, the CP of J1517-2422 ranged between roughly $-0.4\%$ and $0.3\%$ during January--April 2017 and between $-1.0\%$ to $-0.4\%$ between October--December 2017. Given that our $-0.1\%$ detection is well within the historical average and that Sgr A* is at least $10\times$ more circularly polarized than either J1517-2422 or J1744-3116, we robustly detect intrinsic CP from Sgr A*.
            
                \begin{table}
                    \centering
                    \begin{tabular}{|c|c|c|c|}
                        \hline\hline
                         $\nu$ & Stokes I & Stokes V & $p_c$ \\
                         $\rm{[GHz]}$ & [mJy] & [mJy] & [\%] \\
                         \hline
                         \multicolumn{4}{c}{J1517-2422} \\
                         \hline
                         $217.1$ & $3099 \pm 0.54$  & $-3.33 \pm 0.17$ & $-0.11\pm0.005$ \\
                         $220.0$ & $3088 \pm 0.59$  & $-2.37 \pm 0.17$ & $-0.08\pm0.006$\\
                         $233.5$ & $3059 \pm 0.61$  & $-2.75 \pm 0.17$ & $-0.09\pm0.006$\\
                         $235.0$ & $3049 \pm 0.56$  & $-2.16 \pm 0.14$ & $-0.07\pm0.005$\\
                         \hline
                         \multicolumn{4}{c}{J1744-3116}\\
                          & & & \hspace{-2.0cm}$3\sigma$ Upper Limits\\
                         \hline
                         $217.1$ & $ 217.4 \pm 0.17$ & $<\left|0.27\right|$ & $<|0.12|$ \\
                         $220.0$ & $ 214.3 \pm 0.29$ & $<|0.51|$ & $<|0.24|$ \\
                         $233.5$ & $210.2\pm0.15$ & $<|0.27|$ & $<|0.13|$ \\
                         $235.0$ & $207.8\pm 0.17$ & $<|0.27|$ & $<|0.13|$ \\
                         \hline
                    \end{tabular}
                    \caption{The measured Stokes I and V properties for the two non-instrumental polarization calibrators on 16 July 2017. The errors quoted are only statistical. We do not detect CP in J1744-3116 but do detect it in J1517-2422 at a level $\approx-0.1\%$.}
                    \label{tab:calibrator_pc}
                \end{table}
            
            The final aspect to consider is a time-dependent Stokes V leakage. We cannot directly check for this as J1744-3116 is not circularly polarized, and no other gain calibrators were observed. \citet[][Appendix G]{Goddi2021} study the measured Stokes V as a function of feed angle (parallactic angle + receiver rotation relative to the antenna mount) in search for uncalibrated Stokes V terms. In their April 2017 data (near 230 GHz), they found a modulating $\approx0.1\%$ leakage in Stokes V for the nearest calibrators to Sgr A* (J1733-1304 and J1924-2914). This modulation occurs over a range of $\gtrsim 100^\circ$ in the feed angle. In this observation, the feed angle changes by only $\approx8^\circ$. By our estimates, this induces a maximum of $\approx2$ mJy (absolute) variation in Stokes V. As Sgr A*'s Stokes V light curves vary by $\approx15$ mJy, these time-dependent variations are intrinsic to Sgr A* and are not caused by uncalibrated polarization terms.
            

\section{Modeling the Light Curves}\label{sec:analysis}
    We adopt a two-component model consistent with previous work to account for the variable and (quasi-)quiescent components of Sgr A*'s light curves. In contrast to previous work, however, we incorporate a full-Stokes picture. The flaring component is modeled as a homogeneous, spherical synchrotron hotspot adiabatically expanding at a constant speed on a roughly one-hour timescale. This model is characterized by several physical parameters, such as the initial radius, expansion speed, magnetic field strength and orientation, and power-law population of relativistic electrons. Our model does not intrinsically include orbital motion (i.e., a varying magnetic field orientation), gravitational effects (i.e., lensing), non-symmetric geometric evolution (i.e., shearing), nor a sense of the hotspot's location in the accretion flow. We account for secular variations in the accretion flow by modeling the slowly-varying frequency-dependent quiescent component. At each frequency, the four Stokes parameters are assumed to rise or fall linearly during the observation and are characterized by phenomenological parameters, such as gradients with respect to time and reference flux densities. Additionally, the Stokes parameters are frequency-dependent, accounting for physical properties of the accretion flow (such as rotation measure, RM), which we model with spectral indices and gradients with respect to frequency. The two components are described in detail below.
    
    \subsection{Polarization Model for Flaring Emission}
        The Stokes I temporal- and frequency-dependent flaring emission are well-modeled by an adiabatically-expanding synchrotron plasma \citep[thenceforth referred to as a ``hotspot;''][]{VDL1966, FYZ2006}. The hotspot is homogeneous and characterized by five parameters: $I_p$, $p$, $v_{\rm{exp}}$, $R_0$, and $t_0$. $I_p$ is the peak flare flux density at frequency $\nu_0$ at time $t_0$ having radius $R_0$, $p$ is the electron energy power-law index $\left(N(E)\propto E^{-p}\right)$ valid between energies $E_{\rm{min}}$ and $E_{\rm{max}}$, and $v_{\rm{exp}}$ is the (normalized) radial expansion velocity. The flux density at any frequency and size is calculated via
                \begin{align}
                    I_{f}\left(R\right) = I_p\left(\dfrac{\nu}{\nu_0}\right)^{5/2}\left(\dfrac{R}{R_0}\right)^3\dfrac{f(\tau_{\nu})}{f(\tau_0)}.
                \end{align}
        $f(...)$ is a non-trivial function encompassing the full-Stokes radiative transfer equations (briefly described below). $\tau_\nu$, the frequency- and size-dependent optical depth, is given by,
                \begin{align}
                    \tau_\nu(R) = \tau_0\left(\dfrac{\nu}{\nu_0}\right)^{-(p+4)/2}\left(\dfrac{R}{R_0}\right)^{-(2p+3)}.
                \end{align}
        $\tau_0$ is the critical optical depth where the hotspot becomes optically thin and is determined by
                \begin{align}
                    e^{\tau_0} - \left(\dfrac{2p}{3}+1\right)\tau_0-1=0.
                \end{align}
        Finally, we assume a uniform expansion to relate the time and radius:
                \begin{align}
                    R(t) = R_0\left(1 + v_{\rm{exp}}\left(t - t_0\right)\right).
                \end{align}
        Assuming magnetic equipartition, we can determine the hotspot's physical radius and expansion velocity, mass, magnetic field strength, and electron number density.
        
        Given the remarkable high-sensitivity light curves found in Section \ref{sec:data}, we model them using a full-Stokes adiabatically-expanding synchrotron hotspot. To do so, we use the prescription given in \cite{Jones1977a}, which describes the transfer of full-Stokes synchrotron radiation for a static source through a homogeneous medium, i.e., where $v_{\rm{exp}}\ll c$. The Stokes I model described above temporally evolves depending upon the size of the emitting region. Since this temporal evolution is secular, we can model the expanding source as a sequence of hotspots with varying parameters (such as radius, magnetic field strength, and electron population) to account for the time-dependent properties as it grows. Therefore, we convert a stationary solution into a dynamic one to produce full-Stokes light curves of such a source.
    
\begin{figure*}
            \centering
            \includegraphics[width=0.9\textwidth]{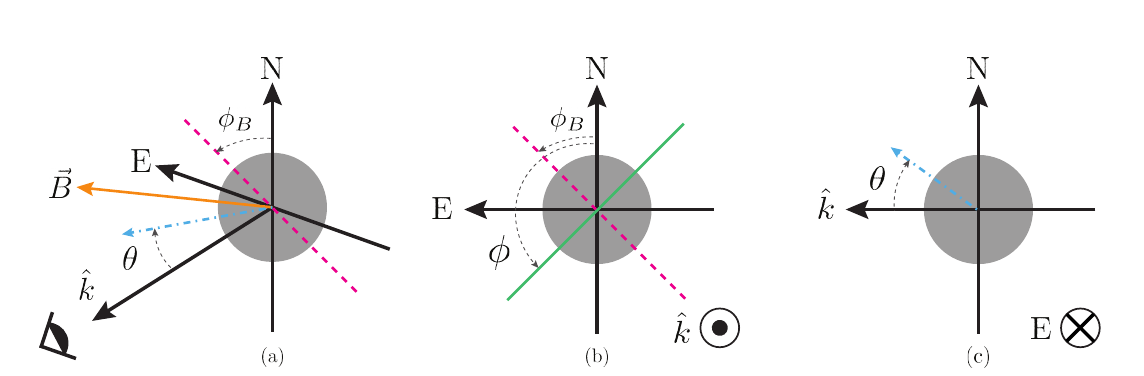}
            \caption{We show several perspectives of the various angles used in this analysis. (a) The general schematic setup, where ``N,'' ``E,'' and ``$\hat{k}$'' denote north, east (in equatorial coordinates), and the unit vector toward the observer, respectively. The hotspot (gray sphere) possesses a three-dimensional magnetic field vector ($\vec{B}$, orange). The pink dashed line denotes the projected magnetic field orientation ($\phi_B$) in the North-East plane perpendicular to the LOS, measured East of North. The dot-dashed cyan arrow shows the angle between the projected magnetic field vector to the LOS ($\theta$). (b) The schematic along the observer's LOS. The dashed pink line again shows the projected magnetic field orientation. In this analysis, we focus on the electric vector position angle (EVPA, $\phi$) shown as a solid green line. The EVPA is also measured East of North. $\phi_B$ and $\phi$ are related by $\phi_B = \phi+\pi/2$ and is wrapped through the $\pi$-ambiguity. For clarity, we do not show the dot-dashed cyan vector. (c) A ``side'' view along the eastern direction to show the projected magnetic field vector along the LOS. Again, for clarity, we do not show the dashed pink or solid green lines.}
            \label{fig:angles}
        \end{figure*}

        Supplementing this model with polarization adds only two parameters: $\theta$ and $\phi$, which are related to the orientation of the magnetic field. A schematic of these angles is shown in Figure \ref{fig:angles}. $\phi$ is the intrinsic electric vector position angle (EVPA; see Equation \ref{eq:polangle}) of the hotspot, measured East of North, projected in the plane-of-sky (POS). It is closely related to the projected magnetic field orientation in the POS, $\phi_B = \phi + \pi/2$. $\phi$ and $\phi_B$ are pseudovectors and obey the $\pi$-ambiguity. $\theta$ is the projected magnetic field vector to the line of sight (LOS). In this convention, $\theta=0^\circ$ and $\theta=90^\circ$ occur when the projected magnetic fields are along and perpendicular to the LOS, respectively.
        
        Under the assumption of homogeneity, the radiative transfer equations as given by \cite{Jones1977a} read
            \begin{align}
             \normalsize{
             \begin{bmatrix}
                    \vspace{0.15cm}
                    J_\nu\\\vspace{0.15cm}
                    \epsilon_QJ_\nu\\\vspace{0.15cm}
                    0\\\vspace{0.15cm}
                    \epsilon_VJ_\nu
                \end{bmatrix}=
                \begin{bmatrix}
                    \left(\frac{d}{d\tau_\nu} + 1\right) & \zeta_Q & 0 & \zeta_V\\
                    \zeta_Q & \left(\frac{d}{d\tau_\nu} + 1\right) & \zeta_V^* & 0\\
                    0 & -\zeta_V^* & \left(\frac{d}{d\tau_\nu} + 1\right) & \zeta_Q^*\\
                    \zeta_V & 0 & -\zeta_Q^* & \left(\frac{d}{d\tau_\nu} + 1\right)\\
                \end{bmatrix}  
                \begin{bmatrix}
                    \vspace{0.15cm}
                    I_f/\Omega\\\vspace{0.15cm}
                    Q_f/\Omega\\\vspace{0.15cm}
                    U_f/\Omega\\\vspace{0.15cm}
                    V_f/\Omega
                \end{bmatrix}}.
                \label{eq:poltransfer}
            \end{align}
        Inside the dielectric tensor, $\tau_\nu$ is the optical depth, $\zeta_{\{Q,V\}}$ are the Stokes Q and V absorption coefficients, $\zeta^{*}_{V}$ is the rotativity coefficient (responsible for Faraday rotation), and $\zeta^{*}_Q$ is the convertability \citep[or repolarization;][]{Pacholczyk1973} coefficient between LP and CP. $\epsilon_{\{Q, V\}}$ are the Stokes Q and V emissivity coefficients, and $J_\nu$ is a source function. $\Omega$ is the solid angle subtended by the source given by $\Omega\equiv\pi R^2/d^2$, where $R$ and $d$ are the radius of the hotspot and its distance from the Earth, respectively. \cite{Jones1977a} provide analytic solutions for the emergent Stokes flux densities ($I_f$, $Q_f$, $U_f$, and $V_f$) integrated over the homogeneous source (see their equations C4-C17). 

        We note the number of assumptions made in this picture, specifically that the hotspot undergoes only secular evolution, which is a limitation in our modeling and differs from other approaches \citep[such as][]{Tiede2020, Gelles2021} that include the hotspot's evolution as it orbits Sgr A*. Our attempt to formally fit this data is to describe the general nature of the expanding hotspot. However, if these secondary processes dominate adiabatic expansion, we would not have expected this modeling to be successful. This is due to the frequency- and polarization-dependent coupling of the polarized radiative transfer equations (Equation \ref{eq:poltransfer} and see Section \ref{ssec:num_fit_params}). These secondary effects can be included and are planned for future work.
        
    \subsection{Quiescent Frequency and Temporal Variations}
        The quiescent component is known to have frequency- and time-dependent baselines that must be accounted for while modeling the flaring emission. The time dependence likely arises from continual, longer-term variability within the accretion flow. For example, \cite{Dexter2014} find an $\sim8$-hour characteristic timescale in Sgr A*'s submm light curves. While their analysis focuses only on Stokes I, we include time-dependent terms for the other three Stokes parameters for consistency. If there is no time dependence in the Stokes Q, U, and V light curves, we expect their time-dependent fitting parameters to be consistent with $0$. Frequency-dependent variations emerge from processes like optical depth (Stokes I and V) or RM (Stokes Q and U).
        
        To account for the frequency- and time-dependent nature of the quiescent component, we use the following model:
            \begin{align}
                I_q(\nu, t) &= \left(I_0 + I_1 \left(t - t_0\right)\right)\left(\dfrac{\nu}{\nu_0}\right)^{\alpha_I}\,,\label{eq:I_quiescent}\\
                Q_q(\nu, t) &= Q_0 + Q_1 \left(t - t_0\right) + Q_2 \left(\nu - \nu_0\right)\,,\label{eq:Q_quiescent}\\
                U_q(\nu, t) &= U_0 + U_1 \left(t - t_0\right) + U_2 \left(\nu - \nu_0\right)\,,\label{eq:U_quiescent} \\
                V_q(\nu, t) &= \left(V_0 + V_1 \left(t - t_0\right)\right)\left(\dfrac{\nu}{\nu_0}\right)^{\alpha_V}\,.\label{eq:V_quiescent}
            \end{align}
        We choose two different frequency dependencies based on the Stokes parameter. The signs of Stokes I and V cannot or do not, respectively, change across the 18 GHz of bandwidth. \citep[The sign of Stokes V does not change from $1.4-340$ GHz;][]{Munoz2012}. Therefore, we use the classic frequency power-law form. Equations \ref{eq:I_quiescent} and \ref{eq:V_quiescent} follow the time- and frequency-dependent model in \cite{Michail2021b}. Due to Faraday rotation, the signs for Stokes Q and U change owing to RM across the bandpass. Therefore, we model frequency-dependent changes in Stokes Q and U using a linear form. Here, $I_i, Q_i, U_i,$ and $V_i$ are all constants; parameters with a ``$0$'' subscript reflect reference flux densities for the four Stokes parameters at frequency $\nu_0$ at time $t_0$. The time- and frequency-dependent slopes are denoted by parameters with subscripts ``$1$'' and ``$2$'', respectively. The spectral indices for Stokes I and V are $\alpha_I$ and $\alpha_V$, respectively. 
        
    \subsection{Results of Model Fitting}\label{ssec:model_fitting}
\begin{figure*}
            \centering
            \includegraphics[trim = 0.8cm 1.55cm 2.25cm 0.5cm, clip, width=0.49\textwidth]{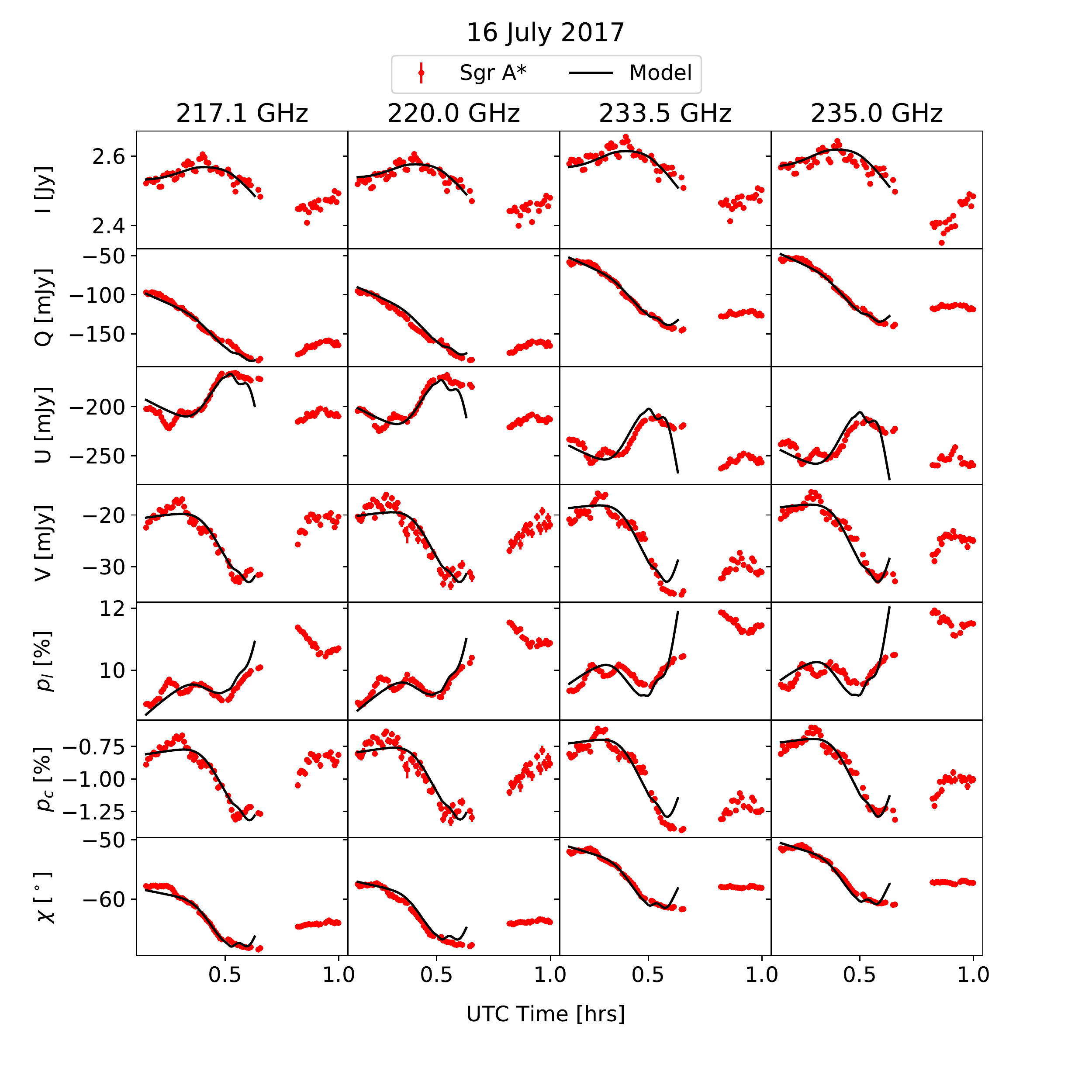}
            \includegraphics[trim = 0.8cm 1.55cm 2.5cm 0.5cm, clip, width=0.49\textwidth]{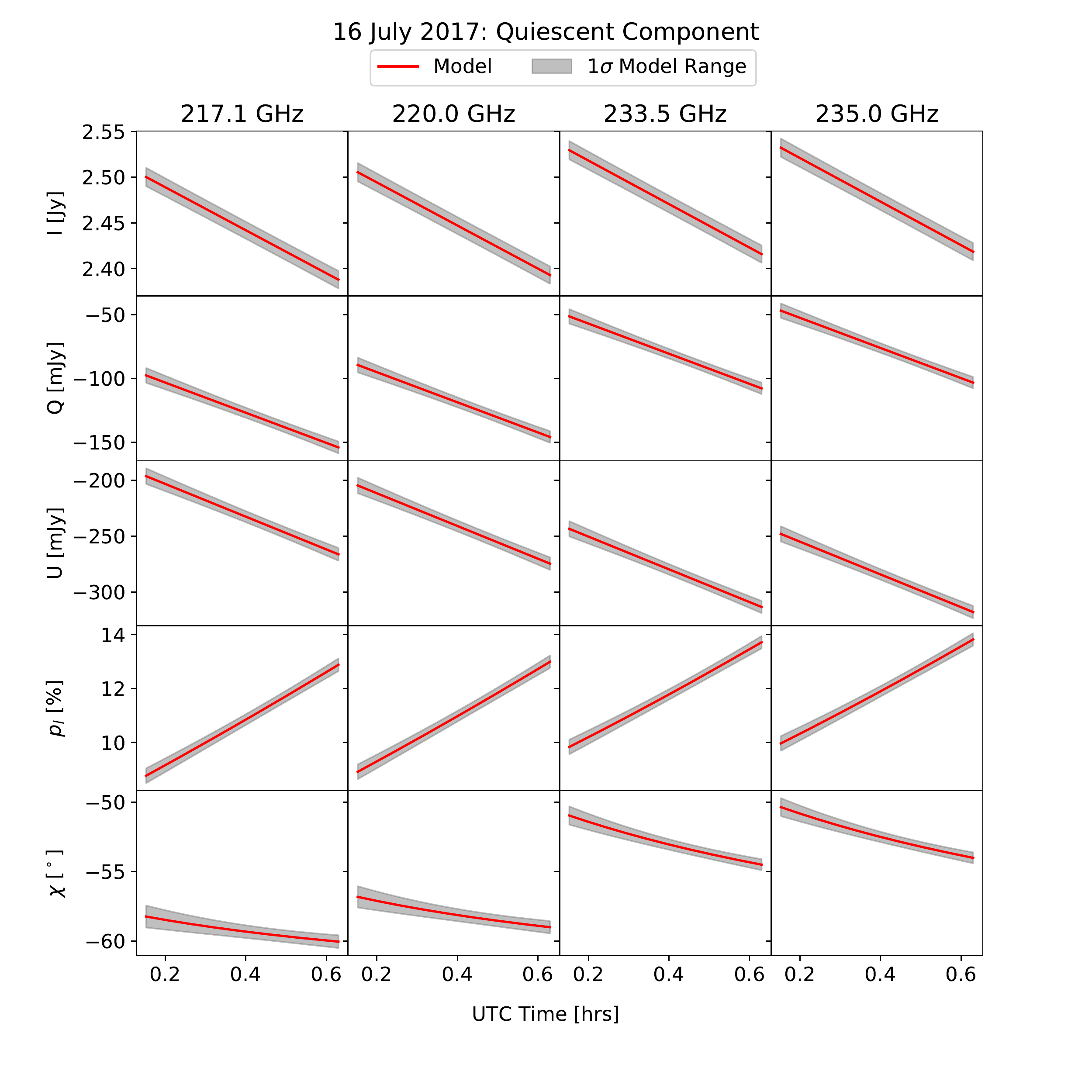}
            \caption{\textit{Left}: Sgr A*'s light curves are shown in red, and the best-fit model is superimposed in black. Due to the short time coverage, we only model the light curves before 00:36 hr UTC. \textit{Right}: Light curves and linear polarimetric quantities for the quiescent component. The best-fit model is plotted in red, and the $1\sigma$ error range is shaded in gray. Unlike the left panel, these figures are plotted to 00:36 hr UTC.} 
            \label{fig:data_model_fits}
        \end{figure*}

        We use \texttt{LMFIT} \citep{lmfit} to simultaneously fit the 16 light curves (4 spectral windows $\times$ 4 Stokes parameters) by minimizing the $\chi^2$ of the variable + quiescent models (i.e., $I_\nu = I_f + I_q$, $Q_\nu = Q_f + Q_q$, $U_\nu = U_f + U_q$, $V_\nu = V_f + V_q$) discussed above. Due to the lack of time coverage, we only fit the data through 00:36 hrs UTC. While there appears to be a second flare beginning near 00:45 hrs UTC, the time coverage is insufficient to model it. Therefore, we do not include those data in the fit. We discuss the implications of this limited time coverage in Section \ref{ssec:caveats}. For this analysis, we set the reference frequency to $\nu_0 = 235.1$ GHz. In Table \ref{tab:fitted_parameters}, we present the fitted parameters values and errors. In Figure \ref{fig:data_model_fits} (left), we show the best-fit model superimposed on Sgr A*'s light curves. 
        
        \begin{table*}
            \centering
            \begin{tabular}{|c|l|c|c|}
                 \hline\hline
                 Parameter & Description & Value & Unit \\\hline
                 \multicolumn{4}{c}{Hotspot}\\\hline
                 $I_p$ & Peak Flare Flux Density at 235.1 GHz & $0.19 \pm 0.01$ & Jy \\
                 $p$   & Electron Power-Law Index     & $3.10\pm0.05$ & --\\
                 $v_{\rm{exp}}$ & Relative Expansion Speed & $1.48 \pm 0.05$ & hr$^{-1}$\\
                 $\phi$   & Intrinsic EVPA projected in POS (East of North) & $144.8 \pm 1.5$ & degrees\\
                 $\theta$ & Angle of projected magnetic field vector relative to LOS & $90.09 \pm 0.01$ & degrees \\
                 $t_0$    & Time of Peak Flux at 235.1 GHz & $0.455 \pm 0.002$ & hr UTC\\
                 \hline
                 \multicolumn{4}{c}{Quiescent Component}\\\hline
                 $I_0$ & Stokes I flux density at $t=t_0$ at 235.1 GHz & $2.46 \pm 0.01$ & Jy\\
                 $Q_0$ & Stokes Q flux density at $t=t_0$ at 235.1 GHz & $-82.6 \pm 3.7$ & mJy \\
                 $U_0$ & Stokes U flux density at $t=t_0$ at 235.1 GHz & $-292.3 \pm 4.7$ & mJy \\
                 $V_0$ & Stokes V flux density at $t=t_0$ at 235.1 GHz & $-16.5 \pm 1.2$ & mJy \\
                 \\
                 $I_1$ & Stokes I Time-Dependent Slope & $-238.2 \pm 13.0$ & mJy hr$^{-1}$\\
                 $Q_1$ & Stokes Q Time-Dependent Slope & $-118.3 \pm 14.2$ & mJy hr$^{-1}$\\
                 $U_1$ & Stokes U Time-Dependent Slope & $-146.6 \pm 16.5$ & mJy hr$^{-1}$\\
                 $V_1$ & Stokes V Time-Dependent Slope & $6.5 \pm 5.2$ & mJy hr$^{-1}$\\
                 \\
                 $\alpha_I$ & Stokes I Spectral Index & $0.16 \pm 0.01$ & --\\
                 $Q_2$ & Stokes Q Frequency-Dependent Slope & $2.82 \pm 0.07$ & mJy GHz$^{-1}$\\
                 $U_2$ & Stokes U Frequency-Dependent Slope & $-2.88 \pm 0.09$ & mJy GHz$^{-1}$\\
                 $\alpha_V$ & Stokes V Spectral Index & $-1.31 \pm 0.45$ & --\\\hline
            \end{tabular}
            \caption{Fitted parameters (with errors) for the joint quiescent and variable model discussed in Section \ref{sec:analysis}.}
            \label{tab:fitted_parameters}
        \end{table*}
        
        \begin{table}
            \centering
            \begin{tabular}{l|l|r}
                \hline\hline
                 Parameter & Description & Value \\\hline
                 \multicolumn{3}{c}{Adopted Parameters}\\\hline
                 $E_{\rm{min}}$ & Electron Lower Energy Bound & $1$ MeV\\
                 $E_{\rm{max}}$ & Electron Upper Energy Bound & $500$ MeV\\\hline
                \multicolumn{3}{c}{Derived Parameters}\\\hline
                 $n_e$ & Electron density & $6.5\times10^{7}$ cm$^{-3}$\\
                 $R_0$ & Radius of flaring region at $t=t_0$ & $9.2\times10^{11}$ cm\\
                 $v_{\rm{exp}}\cdot R_0$ & Physical Expansion Speed & $0.013$c\\
                 $B_{\rm{eq}}$ & Equipartition magnetic field strength & $71$ G\\
                 $M$ & Mass of flaring region & $3.62\times10^{20}$ g\\\hline
            \end{tabular}
            \caption{Adopted and derived hotspot properties from the variable parameters in Table \ref{tab:fitted_parameters}. We do not account for non-relativistic electrons or protons while estimating the equipartition magnetic field strength. Therefore, this is a lower limit on the true value.}
            \label{tab:Plasmon_phys}
        \end{table}
        
        \subsubsection{Variable Component}\label{sssec:var}
            Modeling the light curves gives the six variable component parameters, which characterize the hotspot and are listed in Table \ref{tab:fitted_parameters}. To determine physical parameters, we assume the hotspot is in magnetic equipartition with the electrons responsible for the synchrotron emission between energies $E_{\rm{min}}$ and $E_{\rm{max}}$. In Table \ref{tab:Plasmon_phys}, we present the physical properties of the hotspot fixing $E_{\rm{min}}$ and $E_{\rm{max}}$ to $1$ and $500$ MeV ($\gamma_e\sim2-1000$), respectively. We disregard contributions from protons and non-relativistic electrons in the magnetic field strength, so this is a lower limit on the true value. Overall, we find a $235.1$ GHz peak flare flux density of $0.19$ Jy produced by an electron energy spectrum $N(E)\propto E^{-3.1}$. The hotspot expands at speed $\approx0.013c$ with an equipartition magnetic field strength $71$ G and radius $0.75~R_{\rm{S}}$ ($1~R_{\rm{S}}=1.23\times10^{12}$ cm for a $4.152\times10^{6}~M_{\odot}$ Schwarzschild black hole). Our model robustly detects the two new parameters in this full-polarization fit: $\theta$ and $\phi$. For the intrinsic EVPA of the source, we find $\phi\approx145^\circ$, corresponding to $\phi_B=55^\circ$ East of North ($\phi+\pi/2$ wrapped through the $\pi$-ambiguity). Additionally, we determine $\theta=90.09^\circ$, placing the projected magnetic field orientation approximately perpendicular to the LOS. 
            
            To compare the overall variability of the flaring component to the quiescent emission, we calculate the hotspot's mean LP and CP and their relative fractional change ($\rm{RFC}\equiv \left(\rm{max} - \rm{min}) / \rm{average}\right)$). During the modeled range, we find the flare to have average LP and CP of $\approx35\%$ and $\approx-4.2\%$, respectively, at $235.1$ GHz. The LP goes from a minimum of $\approx9.5\%$ to a maximum of $\approx81\%$, giving an RFC $=~2.04$. The CP ranges from $\approx-15\%$ to $\approx-0.1\%$ with an RFC $=~3.48$.
        
        \subsubsection{Quiescent Component}\label{sssec:quiescent}
            We find statistically-significant time dependencies in the quiescent component's Stokes I, Q, and U light curves. In Figure \ref{fig:data_model_fits} (right), we present the quiescent-only full-Stokes light curves during our modeled range. While the Stokes I time-dependence has been observed previously \citep[e.g.,][]{Michail2021b}, this is the first detection of the quiescent component's Stokes Q and U time-variability. We do not find any changes in the quiescent emission's Stokes V properties, as the time-dependent term is not significant. An uncalibrated Stokes V polarization term (Section \ref{ssec:circular_pol}) would contribute to the final fitted value, further proof that the variations in Stokes V are intrinsic to Sgr A*'s flaring emission. 
            
            We find the quiescent emission has average LP $\approx12\%$ and average CP of $\approx-0.7\%$. The LP ranges between $\approx9.9\%$ to $\approx14\%$, giving RFC $=~0.31$. Since we conclude above the Stokes V quiescent emission is not time-dependent, we do not calculate its RFC. 
            
            Additionally, we identify strong frequency-dependent terms in Stokes I, Q, and U, and a marginal dependence in Stokes V. We find the quiescent emission's Stokes I spectral index is $\alpha_I = 0.15$. The Stokes V spectral index value is much steeper at $\alpha_V = -1.31$. The detection of frequency-dependent slopes in Stokes Q and U generates a non-zero RM in the quiescent emission, which has been found in previous analyses \citep[e.g.,][]{Marrone2006, Bower2018}.
         
\section{Discussion}\label{sec:disc}
    In the previous section, we presented the first-ever full-Stokes modeling of Sgr A*'s total intensity and polarized light curves to simultaneously characterize the quiescent and variable emission. Determining the fitted parameters for our two-component model allows us to study and derive additional physical properties of both components. We derived a few of these properties for the variable emission above by assuming magnetic equipartition. In this section, we compare our results for each component to previous analyses and broadly find them consistent with those in the literature. Finally, given our $\sim40$ minute observation and the 18 free parameters in this fit, we examine their implications on our results.
    
    \subsection{Variable Component}\label{ssec:variable_disc}
        The electron power-law index responsible for the flaring emission is consistent with multi-wavelength constrained values, which broadly range from $p\approx1-3$ \citep[e.g.,][]{FYZ2006,FYZ2008,Eckart2009,Michail2021c,Gravity2021,Witzel2021}. The calculated magnetic field strength is on the higher side of those previously reported, which typically averages a few to tens of Gauss. However, \cite{FYZ2008} and \cite{Eckart2009} report magnetic field strengths $\approx80$ G. These are somewhat stronger than the average field strengths of stellar wind-fed simulations \citep[e.g.,][]{Ressler2020}. It suggests the hotspot may have occurred near the inner accretion flow, where field strengths are stronger and/or in a concentration of flux when the acceleration of particles drives the flaring. The $\approx7\times10^7$ cm$^{-3}$ electron density is consistent with the \cite{Witzel2021} joint variability analysis of Sgr A* at submm, infrared, and X-ray frequencies. 
        
        Previous observations \citep[i.e.,][]{Marrone2007, Gravity2018, Wielgus2022b} have detected variability in the LP angle ($\chi$) of Sgr A* caused by orbital motion. $\chi$ changes by $180^\circ$ over half of the orbital period of the hotspot \citep{Gravity2018}. We only observe $\Delta\chi\sim15^\circ$ over $\sim40$ minutes, the latter of which roughly corresponds to the period at the innermost stable circular orbit (ISCO) for a non-spinning black hole with the mass of Sgr A* ($P = 31.5$ minutes). If the hotspot was near the ISCO, we expect $\Delta\chi\sim180^\circ$ from orbital motion during our modeling range, which would dominate over changes caused by adiabatic expansion. However, the observed change of $\sim15^\circ$ implies the hotspot is far outside the ISCO, and orbital motion-induced variations in $\chi$ are subdominant to changes from the adiabatic expansion. Therefore, the $71$ G field we derive above, which places us towards the inner accretion flow when compared to \citet{Ressler2020}, is likely overestimated. We discuss the possible cause in Section \ref{ssec:caveats}.
            
         We find the magnetic field angle projected on the POS is $\phi_B=55^\circ$ East of North. As a point probe of the conditions within the accretion flow, these results present the first \textit{direct} detection of the accretion flow's projected magnetic field orientation in the POS. Several analyses suggest this position angle is a favored orientation for the Sgr A* system. Near-infrared polarimetric observations of Sgr A*'s flaring emission over several years find a mean EVPA of approximately $60^\circ$ East of North with a range of about $45^\circ$ \citep{Eckart2006, Meyer2007}. \cite{Eckart2006} speculate this indicates the projected spin axis of a disk around Sgr A*, while \cite{Meyer2007} merely propose this as a preferred orientation for the Sgr A* system. Continuum and spectral observations near 1.5 GHz by \cite{FYZ2020} find a symmetric jet-like structure oriented along the Galactic plane at a position angle $\sim60^\circ$, which they attribute to evidence of a jet/outflow from Sgr A*. A more recent analysis by \cite{Wielgus2022b} uses ALMA linear polarimetry at similar frequencies ($\sim220$ GHz) in the context of an orbiting, non-expanding hotspot. Their analysis again confirms a $\sim60^\circ$ EVPA, which they conclude is the hotspot's projected orbital angular momentum axis. \cite{Wielgus2022b} conclude that the orbital motion of a near-infrared hotspot observed by \cite{Gravity2018} is consistent with their results, as well. In the context of these previous observations, we conclude the magnetic and angular momentum axes of the accretion flow are parallelly-oriented, a key signature of magnetically arrested disks \citep[MAD;][]{Narayan2003}.
            
        Of all 18 parameters required to fit our model, $\theta$ (the angle between the LOS and magnetic field vector) is the most well-constrained. For self-absorbed synchrotron sources, \cite{Jones1977a} show the Stokes V absorption, emission, and rotativity ($\zeta_V$, $\epsilon_V$, and $\zeta_V^*$, respectively) coefficients depend on $\theta$. As $\theta\rightarrow90^\circ$, the variations in Stokes V decrease as CP emission and absorption are suppressed. Additionally, the ``strength'' of internal Faraday rotation within the hotspot decreases as $\theta\rightarrow90^\circ$. The convertibility coefficient ($\zeta_Q^*$) is not a function of $\theta$, causing the process of repolarization to be dominant over Faraday rotation. Repolarization has been suggested as one possible explanation for Sgr A*'s low LP but high CP detections at radio frequencies \citep[e.g.,][]{Bower1999b, Sault1999}. If the magnetic field configuration throughout the accretion flow is uniform and stable in time \citep[as suggested by][]{Munoz2012}, then this result corroborates repolarization as the cause of the CP-only detections in the radio.
        
    \subsection{Quiescent Component}
\begin{figure}
                \centering
                \includegraphics[trim = 0.60cm 0.55cm 0.65cm 0.15cm, clip, width=0.6\textwidth]{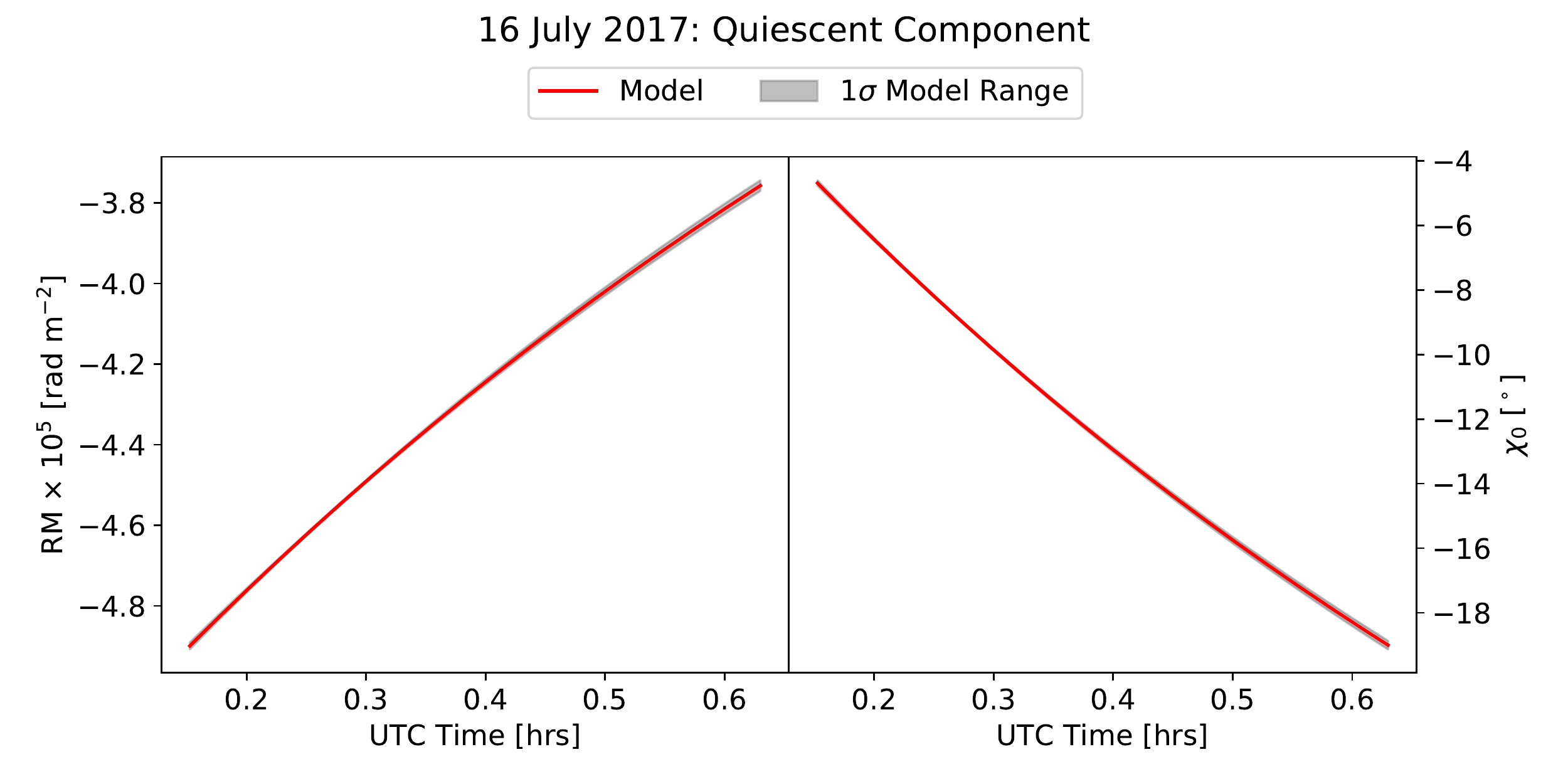}
                \caption{\textit{Left}: A plot of the quiescent component's RM during the modeled time range. Red depicts the best-fit RM value, and the shaded region is the $1\sigma$ model range. \textit{Right}: Similar to the left panel but for the intrinsic EVPA ($\chi_0$) of the quiescent emission.}
                \label{fig:quiescent_pols}
            \end{figure}

        The time- and frequency-dependent nature of the quiescent component is clear and leads us to search for secular variations in the RM and intrinsic EVPA ($\chi_0$). Using an error-weighted linear least squares, we calculate these values by assuming the normal form for Faraday rotation (i.e., $\chi=\rm{RM}\cdot\lambda^2 + \chi_0$). We show the fitted values over the modeled time range in Figure \ref{fig:quiescent_pols}. Overall, we find the RM to vary between $\approx-4.9\times10^{5}$ rad m$^{-2}$ and $\approx-3.8\times10^{5}$ rad m$^{-2}$, while $\chi_0$ ranges between $\approx-4^\circ$ to $\approx-19^\circ$. Variations in these two parameters are strongly anti-correlated, which \cite{Bower2018} also suggest. \cite{Goddi2021} found RMs in the range $-4.84\times10^{5}$ to $-3.28\times10^{5}$ rad m$^{-2}$ and $\chi_0$ ranging from $-18.8^\circ$ to $-14.7^\circ$, which averaged over the entire observation to determine the ``quiescent'' parameters. Our fitted RM and $\chi_0$ match those from their analysis of April 2017 data.
            
        We calculate the average LP and RFC for both components in Sections \ref{sssec:var} and \ref{sssec:quiescent}. We find that the variable emission is $\sim3\times$ more linearly polarized than the quiescent component. Unsurprisingly, the flare's LP properties also vary $\approx6.6\times$ more than the quiescent emission. However, the quiescent emission's modeled LP does change by an appreciable amount (RFC $=~0.31$). Some of its variability may be caused by unmodeled hotspot evolution. The two dominant sources are likely from non-symmetric geometric evolution, such as shearing, and orbital motion. Shearing occurs when the hotspot size is $\sim 0.5~R_{\rm{S}}$ \citep{Wielgus2022b}; in our modeling, we find $R\sim0.75~R_{\rm{S}}$. However, the shearing timescale is similar to the orbital period \citep{Tiede2020}. As discussed in Section \ref{ssec:variable_disc}, the hotspot is not in the inner accretion flow, so the orbital period is longer than our modeling range. The hotspot's orbital motion will affect the measured Stokes Q and U light curves as $\phi$ will vary in the POS and result in some level of hotspot-induced time variability in the quiescent emission's linear polarization properties. These are modelable effects \citep[i.e.,][]{Jones1977b, Wielgus2022b} that will be considered in future work.
        
            
        \cite{Goddi2021} found the Stokes I spectral index ($\alpha_I$) consistent with $0$, whereas we find $0.15$. This is likely explained by a variable spectral index between April and July 2017, which varies on daily to weekly timescales at submm frequencies \citep[see][]{Wielgus2022a}. \cite{Goddi2021} accounted for the $\approx10\%$ absolute uncertainty in ALMA's flux calibration, whereas we only factor in statistical errors. While this tends to underestimate our uncertainty on $\alpha_I$ by $\approx20\%$, it cannot fully account for the discrepancy.
            
        We find $\alpha_V\approx-1.3$, which implies weaker (less negative) Stokes V flux density at higher frequencies. This is in contrast to \cite{Munoz2012} that find Stokes V flux density $\propto\nu^{0.35}$ (more negative) at increasingly higher frequencies. \cite{Bower2018} finds epochs consistent with both positive and negative $\alpha_V$, although they used a linear frequency term instead of a power-law. Despite only having three epochs from which to draw conclusions, there seems to be a general trend in their data. When Sgr A* is brighter, Stokes V is stronger (more negative, $\alpha_V > 0$) at higher frequencies. When Stokes I is lower, Stokes V is weaker ($\alpha_V < 0$) at higher frequencies. In one of three epochs, they find $\alpha_V < 0$ when Sgr A* is $2.68$ Jy at $233$ GHz. Here, we find $\alpha_V<0$ when Sgr A*'s quiescent component is $2.46$ Jy at $235$ GHz. Notably, $\alpha_V > 0$ when Sgr A*'s $227$ GHz flux density was $\approx3.6$ Jy \citep{Munoz2012}. This may suggest a fundamental relationship between Sgr A*'s flux density and the CP spectrum. However, additional data and a more uniform analysis are required before such a correlation is proposed.
            
            \begin{figure}
                \centering
                \includegraphics[width=0.35\columnwidth]{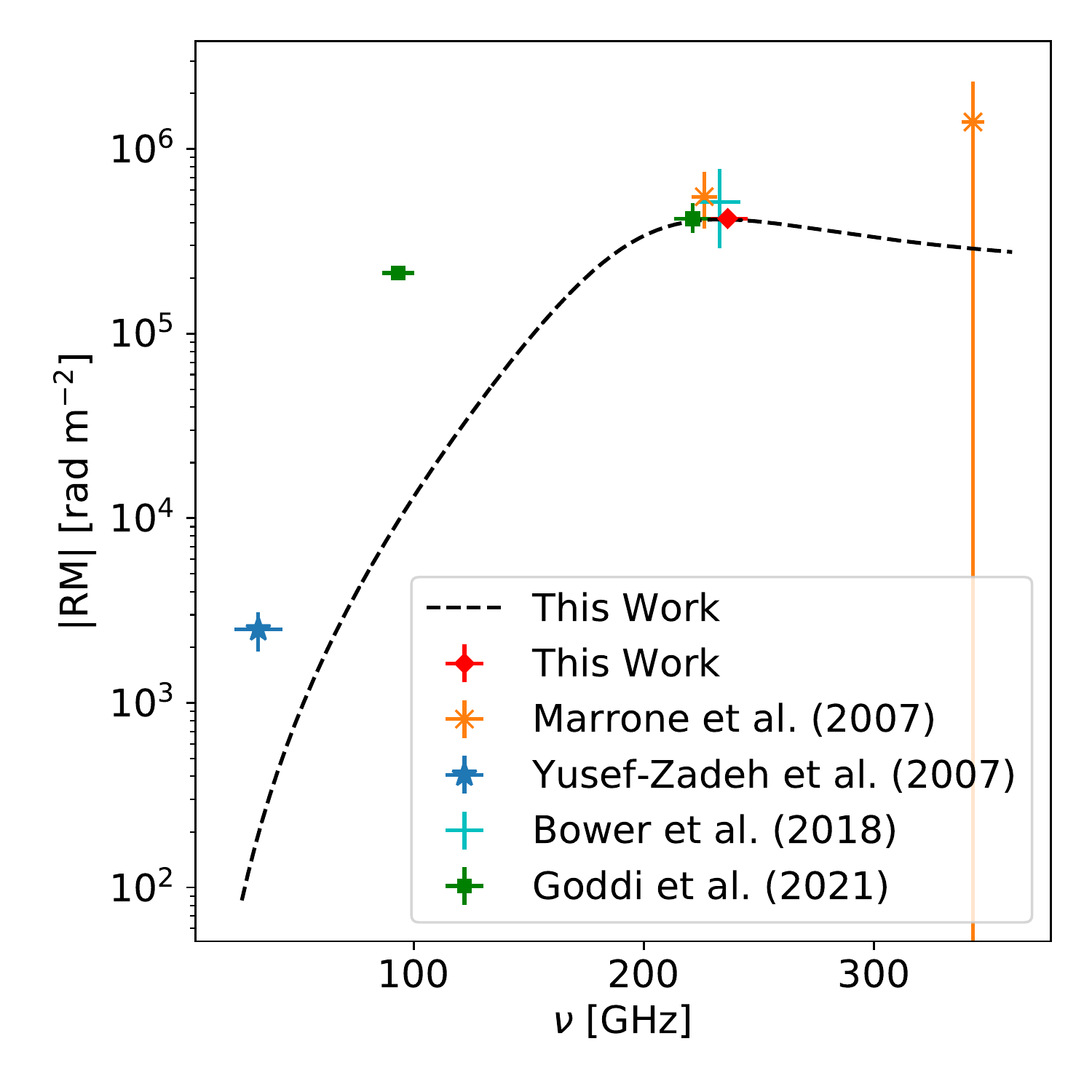}
                \caption{Predicted absolute RM for the quiescent component as a function of frequency from our linear model is plotted as the black dashed line. We compare our predicted RM to those previously published in the literature. \citet[221 GHz]{Goddi2021}, \citet[227/343 GHz]{Marrone2007}, and \citet[233 GHz]{Bower2018} published multiple RM values in a single paper. In these cases, the marker shows the average value while the vertical bars denote the total published range.}
                \label{fig:rm_nu}
            \end{figure}
            
        This phenomenological model predicts a frequency-dependent RM for the quiescent emission. \cite{FYZ2007} also find this trend from compiling published RM values. This suggests that classic Faraday rotation, where the RM is frequency-independent, is not valid for Sgr A*. In Figure \ref{fig:rm_nu}, we plot the predicted (absolute) RM as a function of frequency using our model. We compare these values to those previously published in the literature. At lower frequencies, the model and published values are discrepant by more than an order of magnitude. However, our model predicts the same general trend, where the (absolute) RM falls off at lower frequencies. Unfortunately, the lack of $\approx350$ GHz RM measurements \citep[outside of][]{Marrone2007} makes it difficult to determine if our model underpredicts the average RM, if Sgr A*'s RM significantly changes near this frequency, or a combination of both.
        
    \subsection{Effect of Short Time Coverage}\label{ssec:caveats}
        We only model about half of the light curve on this day (see Section \ref{ssec:model_fitting}). Data after 00:45 hr UTC appear to show the beginning of a new flare. It is impossible to fit this second flare with the hotspot model as the peak was not observed, which is crucial to determine its physical parameters. The short time coverage, compounded with not detecting the beginning of the first flare, complicates fitting the proper quiescent baseline. At submm, the flaring emission is historically $\approx20\%$ of the overall emission. Here, however, $I_p/I_0\approx8\%$. While $I_p/I_0$ can vary, and $I_0$ matches Sgr A*'s flux density in April 2017 \citep{Goddi2021}, this does seem somewhat low. 
        
        We explore the sensitivity to $I_p/I_0$ by assuming $I_p=0.5$ Jy, which is closer to the historic $I_p/I_0\approx20\%$ mentioned above. If $I_p=0.5$ Jy, then the fixed-$I_p$ model has $p=2.53$, $\phi=147.5^\circ$ ($\phi_B=57.5^\circ$), $\theta=90.06^\circ$, and $v_{\rm{exp}}=0.86$ hr$^{-1}$. The hotspot has an equipartition magnetic field strength $55$ G, radius $1~R_{\rm{S}}$, electron density $2.73\times10^{7}$ cm$^{-3}$, and expansion velocity $0.01$c. Noticeably, $\theta$ and $\phi$ are practically unchanged. For completeness, we list other physical properties of the quiescent emission: average RM $=-3.60\times10^5$ rad m$^{-2}$, average $\chi_0=-12.96^\circ$, $\alpha_I=-0.016$, average LP $=18.4\%$, and average CP $=-0.57\%$. While these values are within their nominal ranges, the reduced $\chi^2$ of this fixed-$I_p$ fit is $26\%$ higher than for the model with our best-fitting parameters. We find the properties of the flaring and quiescent components are not extremely sensitive to the value of $I_p$. $\theta$ and $\phi$, which regulate the variations in the Stokes Q, U, and V light curves, are virtually unaffected.
                
    \subsection{Concerning the Number of Fitting Parameters}\label{ssec:num_fit_params}
        We require fitting 18 parameters to model the quiescent and variable components. There is a concern that given the number of parameters, we may be able to fit any data regardless of its true nature. We have taken several steps throughout this analysis as a safeguard, which we detail below.
        
        Of the 18 values fitted here, 12 are dedicated to characterizing the quiescent emission's frequency dependence and temporal evolution. Most of these parameters are phenomenological (i.e., the time-slopes, frequency-slopes, and reference flux densities) as we do not have a physical model for the quiescent emission. To limit over-fitting, we require the quiescent model (Equations \ref{eq:I_quiescent}--\ref{eq:V_quiescent}) to have the same time-slope across all frequencies, but we do not couple these terms across the Stokes parameters. We fit three quiescent terms using four spectral windows for each Stokes parameter. By not pairing terms across Stokes parameters, we guarantee that any non-linear changes in the light curves are due to the variable component.
        
        The final six fitting parameters for the flaring emission characterize these non-linear amplitude variations and link changes in the four Stokes parameters. The resulting full-Stokes light curves have unique patterns depending on the physical parameters of the hotspot. The power of ALMA's wide frequency bandwidths is that we can simultaneously observe Sgr A*'s time and frequency dependence. Since we jointly fit all 16 light curves, variations in any spectral window and/or Stokes parameter must occur in the other light curves following the full-polarization radiative transfer model. Therefore, we conclude that these simultaneous fits cannot model any light curves that do not follow this full-Stokes prescription.
                
\section{Summary}\label{sec:summary}
    We present the first full-Stokes analysis of the adiabatically-expanding synchrotron hotspot model using 230 GHz ALMA light curves of Sgr A* on 16 July 2017. This work is the most robust test of the hotspot model yet performed at any frequency regime by including all four Stokes parameter light curves. The full-polarization modeling we complete is additional proof of the adiabatically-expanding hotspot model, aside from time delay measurements. By modeling the time- and frequency-dependent nature of the variable and quiescent components, we constrain the physical and magnetic properties of the hotspot located within Sgr A*'s accretion flow. Our results are fundamental to the Event Horizon Telescope's future efforts to untangle the full-Stokes variable emission from that of the underlying accretion flow \citep{Broderick2022, EHT2022c, EHT2022d}. Our analysis will benefit from past \citep[i.e.,][]{EHT2022e} and future simultaneous multi-wavelength observations, even those with only total intensity data. These additional data would further constrain the frequency-dependent nature of the variable emission. Our fitted parameters show remarkable consistency with those previously published in the literature. We describe several of our key findings below:
        
    \begin{enumerate}
        \item We observe average LP and CP detections of Sgr A* at levels $\approx10\%$ and $\approx-1\%$ at 235 GHz, respectively, which are consistent with previous measurements.
        \item We find the quiescent component's average RM and $\chi_0$ as $-4.22\times10^5$ rad m$^{-2}$ and $-13.3^\circ$, respectively. These values match well with a recent analysis of Sgr A*'s April 2017 average polarimetric properties \citep{Goddi2021}.
        \item As a point probe of the accretion flow, this hotspot is likely located near the inner edge of the accretion flow owing to the inferred magnetic field strength and electron density being a few times larger than those typically found in MHD simulations \citep[i.e.,][]{Ressler2020}.
        \item The hotspot magnetic field orientation projected on the POS is aligned parallel to the Galactic Plane, matching a previously discovered jet-like feature emanating from Sgr A*, and near-infrared/submm polarization results indicating the accretion flow's angular momentum axis. This reveals the first direct evidence that the accretion flow's magnetic and angular momentum axes are aligned parallel, a key signature of a magnetically-arrested disk.
        \item The hotspot's magnetic field axis is aligned almost perpendicular to the LOS. This suggests repolarization is dominant over Faraday rotation and corroborates it as the cause of low-LP but high-CP in Sgr A* at radio frequencies.
        \item We find the results of the variable component are not drastically (or at all) altered from the lack of long-duration time coverage provided by these data.
    \end{enumerate}
        
    Several exciting prospects remain to test this model. There is a diverse dataset of full-track, long-duration ALMA and Submillimeter Array (SMA) observations of Sgr A*. Many of these are simultaneous observations at similar or vastly different frequencies. In the first case, these data provide full-Stokes light curves over a broader time range than a single array could provide. In the latter case, multiwavelength observations allow us to constrain the frequency-dependent total intensity and polarized nature of the quiescent emission and solidly test this full-Stokes hotspot model across a wide range of frequencies. Future analyses will benefit from fast-frequency switching or sub-array  ALMA observations (for example, simultaneously using the 12-meter and 7-meter arrays at two separate frequencies). This expanded analysis will allow us to test for variability in the hotspot's physical parameters, such as the magnetic field orientation and strength. For example, if $\phi$ is variable, as is suggested in the near-infrared \citep[i.e.,][]{Eckart2006}, its range may signify the opening angle of an outflow emanating from Sgr A*.
        
\section*{Acknowledgements}
    We thank the anonymous referee for their very helpful and constructive comments, which strengthened the arguments and analysis in this work. The National Radio Astronomy Observatory is a facility of the National Science Foundation operated under cooperative agreement by Associated Universities, Inc. This paper makes use of the following ALMA data: ADS/JAO.ALMA\#2016.A.00037.T. ALMA is a partnership of ESO (representing its member states), NSF (USA) and NINS (Japan), together with NRC (Canada), MOST and ASIAA (Taiwan), and KASI (Republic of Korea), in cooperation with the Republic of Chile. The Joint ALMA Observatory is operated by ESO, AUI/NRAO and NAOJ. This research was supported in part through the computational resources and staff contributions provided for the Quest high performance computing facility at Northwestern University which is jointly supported by the Office of the Provost, the Office for Research, and Northwestern University Information Technology.

\section*{Data Availability}
    The data used in this analysis are publicly available from the ALMA archive. The light curves and code used to model these data are available upon request to the first author.

\bibliographystyle{mnras}
\bibliography{almapol} 

\appendix\section{Polarization Conventions}\label{appendix:pol_conv}
    Our code (Section \ref{ssec:observations}) provides the four Stokes light curves (I, Q, U, and V) and their uncertainties ($\sigma'_{\{I,Q,U,V\}}$). The primed uncertainties do not properly account for $\approx0.03\%$ systematic leakage of Stokes I into Stokes Q and U \citep{Goddi2021}. We follow their analysis by defining the ``true'' uncertainties $\left(\sigma_{\{Q,U\}}\right)$ as the quadrature sum of the measurement uncertainties and systematic leakage,
        \begin{align}
            \sigma_{\{Q,U\}} = \sqrt{\sigma'^2_{\{Q,U\}} + (0.0003I)^2}.
        \end{align}
    The measured linear ($p_{m,l}$) and circular ($p_c$) polarization percents are:
        \begin{align}
            p_{m,l} &= 100\% \times \dfrac{\sqrt{Q^2+U^2}}{I},\\
            p_c &= 100\% \times \dfrac{V}{I}.
        \end{align}
    Q and U can take any real value, but $p_{m,l}$ is strictly positive, which positively biases the measured LP. We follow \cite{Serkowski1974} by debiasing this quantity as:
        \begin{align}
            p_l = \sqrt{p^2_{m,l} - \sigma_{p_l}^2},
        \end{align}
    where $\sigma_{p_l}$ is the uncertainty in the LP percent,
        \begin{align}
            \sigma_{p_l} = 100\% \times \dfrac{1}{I}\sqrt{ \dfrac{Q^2\sigma_Q^2 + U^2\sigma_U^2}{Q^2 + U^2} + \dfrac{\sigma_I^2}{I^2}\left(Q^2 + U^2\right)}.
        \end{align}
    The CP percent can be both positive or negative; it is, therefore, unbiased. The uncertainty on $p_c$ is
        \begin{align}
            \sigma_{p_c} = 100\% \times \dfrac{1}{I}\sqrt{\sigma_V^2 + \dfrac{\sigma_I^2}{I^2}V^2}\,.
        \end{align}
    Finally, we define the electric vector polarization angle (EVPA), $\chi$, measured East of North:
        \begin{equation}
            \chi = \dfrac{1}{2}\rm{arctan2}~\left(\dfrac{U}{Q}\right),
            \label{eq:polangle}
        \end{equation}
    where the $\rm{arctan2}$ function places the angle in the correct quadrant. Polarization is a pseudovector causing a $180^\circ$ ambiguity in the LP orientation. Therefore, $\chi$ is defined only on the interval $\left[-90^\circ, 90^\circ\right]$. Finally, we define the uncertainty in the polarization angle, $\sigma_\chi$:
        \begin{align}
            \sigma_\chi = \dfrac{1}{2}\dfrac{1}{Q^2+U^2}\sqrt{Q^2\sigma_U^2 + U^2\sigma_Q^2}\,.
        \end{align}
\bsp	
\label{lastpage}
\end{document}